\documentstyle[12pt,aaspp4]{article}
\input{psfig}
\def\arcsec {$^{\prime \prime}$}
\def\arcmin {$^{\prime}$}

\def\cmmb   {~cm$^{-2}$}

\def\deg    {^{\circ}}
\def\degr   {$^{\circ}$}
\def\etal   {{\it et~al.\/}}
\def\HI     {H~{\sc I}}
\def\HII    {H~{\sc {II}}}
\def\kms    {~km~s$^{-1}$}

\begin{document}

\title{Detection of Cold Atomic Clouds in the Magellanic Bridge}
\author{Henry A. Kobulnicky\footnote{Hubble Fellow}
\footnote{Visiting Astronomer, Australia Telescope National Facility, 
Epping, N.S.W., Australia}}\affil{University of California, Santa Cruz \\ 
Lick Observatory \\ 
Santa Cruz, CA, 95064  \\ 
Electronic Mail: chip@ucolick.org}
\authoremail{chip@ucolick.org}
\author{John M. Dickey$^2$}

\affil{University of Minnesota, Department of Astronomy \\
116 Church St. SE, Minneapolis, MN 55455 \\ Electronic Mail: 
john@astro.spa.umn.edu}
\author{Accepted for Publication in The Astronomical Journal}

\vskip 1.cm

\begin{abstract}

We report a direct detection of cold atomic hydrogen in the Magellanic
Bridge using 21-cm absorption spectroscopy toward the radio source
B0312-770.  With a maximum absorption optical depth of $\tau=0.10$ and
a maximum 21-cm emission brightness temperature of 1.4~K
($N_{HI}=1.2\times10^{20}$ \cmmb), this line of sight yields a spin
temperature, $T_s$, between 20~K and 40~K.  \HI\ 21-cm absorption and
emission spectroscopy toward 7 other low \HI\ column density sightlines
on the periphery of the LMC and SMC reveal absorption  toward one
additional background radio source behind the SMC with $\tau=0.03$.
The data have typical sensitivities of $\sigma_{\tau}=0.005$ to $0.070$
in absorption and $\sigma_{T_B}=0.03$~K in emission.  These data
demonstrate the presence of a cold atomic phase which is probably 
accompanied by molecular condensations in the tenuous interstellar
medium of the Bridge region.  Young OB stars observed in the Magellanic
Bridge could form {\it in situ} from these cold condensations
rather than migrate from regions of active star formation in the main
body of the SMC.  The existence of cold condensations and star
formation in the Magellanic Bridge might be understood as a small scale
version of the mechanism that produces star formation
in the tidal tails of interacting galaxies.

\end{abstract}


\section{Introduction}

As our nearest dwarf galaxy neighbors\footnote{We adopt a distance of
48 kpc to the LMC and 60 kpc to the SMC, so that 1\arcsec\ corresponds
to linear sizes of 0.23 pc and 0.29 pc, respectively,
although the SMC apparently has a large line-of-sight depth
(Wayte 1990)}, the Magellanic
Clouds provide the best laboratories to investigate the physical
properties of the interstellar medium in relatively metal-poor
environments.   The impact of reduced heavy element abundances on the
heating and cooling of the interstellar medium (ISM), cloud collapse,
and star formation is poorly constrained from an observational
standpoint.  Reduced metal abundance produces the lower CO surface
brightness of molecular gas in the LMC and SMC (Israel \etal\ 1993;
Rubio, Lequeux, \& Boulanger 1993), making the molecular phase harder
to trace.   

Measuring the spin temperature, $T_s$, of the atomic medium is one
method of measuring the temperature structure of the ISM.  Mebold \etal\
(1991, 1997), Braun \& Walterbos (1992), 
Dickey \& Brinks (1993), Braun (1997), and
Dickey \etal\ (1994, 1998) have pioneered the study of interstellar
temperatures in the Magellanic Clouds, M~31, and M~33 using this
method.  In the inner disk of M~31, the spin temperature of the cold
neutral component of ISM appears to be similar to that in the solar
neighborhood, but shows a radial gradient, averaging 70~K in the inner
disk and rising to 175~K in the outer disk beyond 10 kpc (Braun \&
Walterbos 1992).  Dickey \etal\ (1994) and Mebold \etal\ (1997) find
that in the LMC the average temperature of the cold clouds is 30-40~K.
While the abundance of cold clouds appears to be concentrated in
regions of strong star formation like 30 Doradus, it is not yet clear
from existing data whether the cold clouds in the LMC have a lower spin
temperature than the Galaxy, or simply a greater volume filling factor.  Contrary to
results in M~31 where the spin temperatures are higher at large radii,
the lowest spin temperatures measured in the LMC occur in low column
density regions on the outskirts of the galaxy.  A lower heating rate
in the outskirts of the LMC might explain the decrease in spin
temperature with radius.

In order to study the impact of environment on the formation of cold
clouds and the physical conditions of the ISM, we have obtained 21-cm
absorption and emission spectra along 8 lines of sight in the low
column density periphery of the LMC and SMC.  Figure~1
shows the locations of the background sources superimposed on
a 21-cm neutral hydrogen peak brightness
map of the Magellanic System (Putman \etal\ 1998).  One 
sightline passes
through the Magellanic Bridge, sampling the tenuous medium between the
Clouds where we detect 21-cm absorption from cool atomic clouds.   In
this paper, we describe the collection and analysis of 21-cm emission
spectra from the Parkes 64 m telescope and interferometric spectroscopy
from the Australia Telescope Compact Array toward background radio
continuum sources.  We discuss the implications for the derived spin
temperatures on the physical conditions in the Magellanic Bridge and
on the potential for star formation in cold atomic gas clouds observed
in the tidal tails of interacting galaxies.

\section{Observations}

\subsection{Parkes Radio Telescope 21-cm Emission Line Spectroscopy}

We used the Parkes 64 m radio telescope with the single-beam receiver
to measure the 21-cm neutral hydrogen emission profiles toward eight
background radio sources in and around the Magellanic Clouds.  The
observations were conducted on 1997 February 12--13, using a total
bandwidth of 8 MHz and 2048 channels of 0.0039 MHz (0.82 \kms) centered
at $v_{\odot}=250$ \kms.  Due to the extended nature of HI emission
around the Magellanic Clouds, all potential reference positions are
contaminated  by \HI\ emission.  To obtain emission-free reference
spectra we interleaved on-band integrations with off-band
frequency-switched ($\Delta{v}$=3 MHz) integrations centered on
$v_\odot=$880 \kms, well beyond the velocity range of Magellanic
\HI\ emission.  In addition to the line of sight toward the background
radio source, we observed two to four adjacent positions, offset by
14\arcmin\ (slightly less than the 15\arcmin\ half-power beamwidth) in
the cardinal directions.  The system temperature averaged 33~K and the
mean RMS noise in the resulting spectra is 0.03~K.  We observed the
Galactic \HII\ region, S8, as a flux calibration standard, and scaled
our temperatures to match the 76~K reported by Kalberla, Mebold, \&
Reif (1982).

The data for left and right circular polarization were reduced by
subtracting the frequency-switched reference spectrum from the source,
and fitting a low order polynomial to emission-free regions of the
spectrum to remove any residual baseline.  Since the Galactic and
Magellanic \HI\ emission features appear both in the signal and
reference spectra, separated by 630 \kms, we are able to invert the
signal and reference spectra, perform the same reduction procedure,
and effectively double the integration time.

The emission spectra toward the eight sightlines appear in
Figures~2a--h.  The largest source of uncertainty in the 21-cm flux and
the shape of emission features associated with the Magellanic Clouds is
due to ambiguities in the baseline subtraction.  Narrow ($<$40 \kms)
emission features brighter than 0.1~K should all be easily detected in
these spectra, while broad ($>$100 \kms) low level features may be
missed or mis-measured due to the baseline fitting procedure.

\subsection{ATCA 21-cm Absorption Line Spectroscopy}

We used the Australia Telescope Compact Array (ATCA)\footnote{The
Australia Telescope is funded by the Commonwealth of Australia for
operation as a National Facility funded by CSIRO} on 1997 January 3 in
the 750A configuration to search for the 21-cm transition in absorption
by cool clouds toward 8 background compact radio sources in Table~1.
The target sources were selected from catalogs of known QSOs (e.g.,
Hewitt \& Burbidge 1993) behind the LMC and SMC which had radio fluxes
of at least 100 mJy in the Parkes radio survey (Otrupcek \& Wright
1991).  We included the V=16 QSO 0242-729 in our observing program
although it did not have a previous radio detection.  We avoided radio
sources observed in previous Magellanic Cloud absorption-line surveys
(Dickey \etal\ 1994; Dickey \etal\ 1998) which are located behind the
main body of the SMC and LMC in regions of higher \HI\ column density.
Our targets have one object in common with the \HI\ absorption survey
of Mebold \etal\ (1991), B0202-765.

The ATCA correlator was configured in the FULL-4-2048 mode, yielding
2048 channels with width 1.95 kHz (0.41 \kms).  During a 13-hour track,
we observed each source for six 15-minute intervals, separated by at
least 1 hour to let the earth's rotation  maximize the UV
coverage.  Once per hour we observed an unresolved phase calibrator,
B0252-712, for five minutes.  At the beginning and end of the track we
observed PKS~1934-638, a standard flux and bandpass calibrator for the
Compact Array.  The data were reduced and calibrated in the standard
manner, using the ATNF implementation of the AIPS software.  The RMS
noise in the bandpass solution derived from the observations of
B1934-638 is 30 mJy, or 0.2\% of the continuum level.

We used the AIPS task IMAGR to transform the UV visibility data into
spatial maps, clean the maps to remove sidelobe structures, and produce
spectral cubes with 0.82 \kms\ channel separation.  The cleaned maps
have a typical RMS of 22 mJy/beam.  In each field, (except B0242-729
where we find no radio continuum source to a 3$\sigma$ limit of 60 mJy) we detect
the background quasar which is by far the strongest radio source in the
primary beam. Table~1 lists the measured flux density for each quasar,
along with the flux measured by Otrupcek \& Wright (1991) in the Parkes
PKSCAT90 survey.  For most of the objects, the two fluxes are in
excellent agreement.  B2300-686 does not appear in the PKSCAT90,
presumably because its flux density, as measured here by the ATCA, is
at the detection limit of that survey.  B0637-752 appears 1.2 Jy
brighter in PKSCAT90 compared to our measurement, plausibly due to the
intrinsic variability of most quasars.

All objects except B2300-683 are consistent with point sources at the
resolution of our maps, typically 40\arcsec.  A Gaussian deconvolution
using the synthesized clean beam suggests an intrinsic size of
60\arcsec\
$\times$ 4\arcsec\ for B2300-683 with a major axis position angle of
118\degr.

We extracted one-dimensional spectra toward each quasar by summing the
4-pixel region of peak flux density in each frequency plane of the
spectral cube.  The resulting spectra appear in Figures~2a--h,
smoothed with a 3-channel boxcar for display purposes.

\section{Spectral Analysis}

\subsection{HI Emission Profiles and Column Densities}

\HI\ emission from the Milky Way appears in all of the Parkes spectra
centered near $v_{\odot}=10$ \kms.  For the remainder of the analysis,
we ignore the Galactic contribution to \HI\ column densities ($v_\odot
< 100 $ \kms ) which is clearly separated from Magellanic Cloud
velocities ($v_\odot > 150 $ \kms ).  Along six sightlines (B0202-765,
B0242-729, B0312-770, B0506-612, B0637-752, and B2353-685) we detect an
\HI\ intensity in excess of 1~K~\kms\ at velocities  between 150 and
350 \kms\ corresponding to the Magellanic system.  Toward 2 sources,
B0743-683 and B2300-683, we achieve only upper limits of
$N_{HI}<0.1\times10^{19} cm^{-2}$ .  In Table~2 we compile the integral
brightness temperature, $I_{HI}$, summed across the velocity range
150---350 \kms\ for each line of sight.  The suffixes n,s,e,w denote
sightlines adjoining the primary targets, offset by 14\arcmin\ in the
equatorial north, south, east, and west directions.  The stated
uncertainties correspond to
$\sqrt{Nchannels}\times\delta{v}\times\sigma_{RMS}$ where $\sigma_{RMS}$
is the RMS of the offline region.  This parameterization does not take
into account potential systematic errors due to baseline subtraction
problems or broad emission components, which, if present, would dominate 
the measurement uncertainties.  Table~2 lists the total
\HI\ column densities along each line of sight due to the Magellanic
Clouds using (Spitzer 1978)

\begin{equation}
N_{HI}=1.823\times10^{18} \int T_B~dv~(cm^{-2}) 
\end{equation}

\noindent for an optically thin medium.  The data in Table~2 show that
the \HI\ column densities vary by up to 50\% at adjoining positions
around the primary line of sight.  This dispersion suggests the
presence of significant column density variations within the primary
beam.  It serves as an important reminder that the large
(15\arcmin) beam of the \HI\ emission data compared to the pencil-beam
nature of the absorption observations necessarily means that the two
measurements refer to different volumes of the ISM and should be
compared in a statistical sense only.

Using the ATCA data, we mapped each field using only UV data from the
shortest baselines (applying a heavy UV taper of 5 kilo-lambda) to
search for clumpy \HI\ emission on scales smaller than the
15\arcmin\ Parkes beam.  Since the interferometer acts as a spatial
filter, the data are insensitive to uniform emission features larger
than a few arcminutes across.  At 21-cm, the shortest baselines in this
configuration of the ATCA are sensitive only to structures with angular
sizes less than $\sim$200\arcsec.  In 7 of the 8 fields, we do not
detect evidence for \HI\ emission on scales smaller than this angular
size, not even in a statistical sense.  The RMS fluctuations in
channels corresponding to velocities with \HI\ emission in the Parkes
spectra are identical to those in off-line channels.  Only in the
B0242-729 field do we find brightness fluctuations in the velocity
interval 194---215 \kms, corresponding to the emission from the single
dish spectra.  The maps show emission from 6-8 knots with deconvolved
dimensions of  $\sim$10 arcsec, or 2-4 pc.  The maximum flux detected
in any single channel is less than 0.27 Jy near 200 \kms.  Assuming a
gain of 0.74~K/Jy for Parkes 64 m telescope, these features would
produce a brightness temperature of 0.20~K while the observed
single-dish brightness temperature exceeds 6~K at this velocity
(Figure~2b).  Thus, only a small faction  ($<0.2/6.2=3$\%) of the
emission arises from compact high surface brightness structures.  The
low incidence of such compact structures suggests that the single dish
(15\arcmin) spectra are an adequate measure of the mean H~I column
density in the direction of the compact radio continuum sources.

\subsection{HI Absorption Spectra}

Toward two sources (B0202-765 and B0312-770), we detect 21-cm
absorption due to cool atomic hydrogen in the lower spin configuration
of the ground $1^2S_{1/2}$ state.  Along both sightlines, the velocity
of the maximum optical depths ($\tau=0.03$ and $\tau=0.10$ respectively)
agrees well with the velocity of maximum brightness temperature, $T_B$,
suggesting that the features arise within the same physical
interstellar structures.  Table~2 contains the equivalent widths of the
absorption features integrated across the velocity range where the
absorption signature exceeds 2$\sigma$.  For the six objects without
observed \HI\ absorption, we achieve $3 \sigma$ upper limits on the
\HI\ optical depth between $\tau<0.22$ and $\tau<0.015$, depending on
the continuum source flux density.  Table~2 contains the RMS optical
depth sensitivity for each of the ATCA spectra.

For the two sources with measurable 21-cm absorption, we perform a more
detailed analysis of the spectra 
by fitting Gaussian components to the emission and
absorption features.  B0312-770 exhibits three clear components in both
emission and absorption.  In Table~2 we list the integrated brightness
temperatures or optical depths, the center velocities, FWHM, and peak
amplitudes for each of the components individually.  We fit all 3
components simultaneously, allowing the position, amplitude, and width
to vary as free parameters.  Attempts to use only two components result
in large residuals, while using four or more components does not
significantly improve the fit.  Of course, many narrow, blended
components may produce the observed spectra, but with the current
spectral resolution and limited signal-to-noise, there is no warrant for
using more than 3 components in the fit.

For B0202-765 two Gaussian components produce an excellent fit to the
emission spectrum.  The dominant component is centered at
$v_\odot=$161.8 \kms\ and a much weaker component appears at
$v_\odot=$197.8 \kms. For the
absorption spectrum, the situation is more complicated.  A single
component fit centered at $v_\odot=$156.1 \kms\ has large residuals.
Adding a second component results in slightly smaller residuals and
yields center velocities of $v_\odot=$146.6 and $v_\odot=$162.7 \kms
for the two features.  Adding a third component significantly improves
the fit.  The three absorption components are centered at
$v_\odot=$137.4 \kms, 147.6 \kms, and  162.1 \kms.  The velocity of the
dominant component at $v_\odot=$162.1 \kms\ agrees well with the
velocity of the main emission component.  One reasonable interpretation is
that the interstellar \HI\ structure producing the dominant emission
component contains three separate cold inclusions at different
velocities which give rise to the three absorption components.  If
there exists cold atomic gas associated with the weaker emission
component at $v_\odot=$197.8 \kms\, it must be below our detection 
threshold of $\sigma_{\tau}=0.01$.

\section{The Physical Conditions of Gas in the Magellanic Bridge}

Measurements of the 21-cm emission and absorption profiles along the
same line of sight provide an estimate of the spin temperature, $T_s$, of
the neutral atomic medium.  However, any given sightline will sample a
mixture of physical components, including the diffuse ionized gas, the
warm and cold neutral medium,  and even the dense molecular phase.   In
the Milky Way, neutral hydrogen exists predominantly (75\%) in the warm
(200~K) neutral phase.  The results in other galaxies are mixed, with
M~31 having 65\% in the warm phase, and M~33 having 85\% (Dickey \&
Brinks 1993).  The results for the LMC are consistent with the Milky
Way, except for the 30-Doradus region which appears to have an excess
of cold atomic clouds (Dickey \etal\ 1994).  The SMC shows 
spin temperatures between 20 and 50~K (Dickey \etal\ 1999).
However, the low column density regions between the Magellanic Clouds
have not yet been probed.  The radio source B0312-770 lies behind the
Magellanic Bridge and offers an opportunity to probe its thermodynamic
state.

Using the data from Table~2, we explore several methods for finding the
spin temperature of the gas, $T_s$.  In the second column of Table~3,
we calculate a ``mean spin temperature'' along each line of sight using

\begin{equation}
<T_s> =  {{{\int{T_B}dv}\over{\int{1-e^{(-\tau)}dv}}}}\simeq 
	{ {{\int{T_B}dv}\over{\int{\tau}dv}} }.
\end{equation}

\noindent This algorithm sums all the gas along the line of sight.  It
does not distinguish  emission from the warm ISM associated with the
cold absorption features from other, unaffiliated warm gas along the
same line of sight.  Thus, it provides an upper limit on the spin
temperature of the cold neutral medium.  We find mean spin temperatures
of 87$\pm$7~K and 32$\pm$4~K for B0202-765 and B0312-770 respectively.
Computing a mean spin temperature for each well-defined
absorption/emission feature in these two sources yields 143$\pm$19~K
for component 3 in B0202-765 and 22---46~K for the three components in
B0312-770.

Another approach uses only the emission brightness temperature and
maximum optical depth of associated emission/absorption components
which co-incide at velocity $v_0$.  In the third column of Table~3 we
list the ``central spin temperature'' defined by

\begin{equation}
T_s(v_0) = {{T_B(v_0)}\over{(1-e^{-\tau(v_0)})}},
\end{equation}

\noindent where $v_0$ is center velocity of each absorption component
in Table~2, and $T_B(v_0)$ is the brightness temperature interpolated
at $v_0$.  The resulting spin temperatures using this approach are
lower than for the line integral algorithm since it is less sensitive
to emitting gas at different velocities which is presumably not
affiliated with the absorbing component.  However, if different warm
and cold gas components overlap in velocity space, then even this
method measures only the ensemble average of the physical
properties.    Using this approach, we find spin temperatures of 8~K,
38~K, and 68~K for the three components toward B0202-765 and 6~K, 38~K,
and 14~K toward B0312-770.

Mebold \etal\ (1997) develop a more sophisticated method for measuring
the spin temperature of the ISM by separating the warm and cold
components.  Cold, absorbing gas clouds are usually embedded amidst
more extended complexes of warm neutral gas responsible for the
emission.   An accurate measurement of the spin temperature requires
separating the emission affiliated with the absorbing components from
emission produced by more extended, unrelated gas.   In Figures~3 and 4
we plot the emission brightness temperature, $T_B$, versus the
absorption ($1-e^{-\tau}\sim\tau$ for the small optical depths studied
here) on a per-channel basis.  We compute the emission brightness
temperature at the velocity of the absorption velocity channels using a
cubic spline interpolation.  A cross marks each channel.  A dotted line
connects channels with emission brightness temperature greater than
2$\sigma$, while a solid line connects four or more consecutive
channels with absorption at greater than the 2$\sigma$ level.  Radial
lines from the origin denote the locii of constant spin temperature.
Parcels of warm gas without absorbing components cause the sequence of
connected points to make an excursion in the vertical direction.
Absorbing gas clouds with measurable optical depth cause the series of
connected points to make an excursion to the right (solid lines).  In
the method outlined by Mebold \etal\, the slope, $m$, of a linear fit
to the absorbing channels outlined by the solid lines yields the spin
temperature of the cold component alone, $T_{sc}$.  The intercept of
this linear fit with the y-axis yields the emission brightness
temperature, $T_{EW}$, of the unaffiliated warm gas.  Our attempts to
measure the spin temperature of the cold component alone using this
method yield unphysically low (even negative) values.  The unphysical
results are probably because large beamsize of the emission
observations is poorly matched to the scale of the individual absorbing
material, and there is only a statistical correlation between
absorption and emission properties of any particular velocity channel.
Consequently, we proceed no further with this type of detailed analysis

\subsection{Comments on Individual Objects}

{\it B0202-765:\ } Mebold \etal\ (1991) presented ATCA spectroscopy toward
this source, and they detected no absorption to a level of
$\tau>0.044$.  Our new data have four times the sensitivity of that
presented in Mebold \etal.  We detect absorption with a maximum
optical depth of $\tau=0.025$ in the strongest component.

{\it B0242-729:\ } Although this sightline contained no known radio source
in previous single-dish radio surveys, we included this field because
it contains an optically bright background QSO.  We find no radio continuum source
to a 3$\sigma$ limit of 66 mJy.

{\it B0312-770:\ } Situated behind the Magellanic Bridge, this source
allows us to probe the physical conditions in the tenuous Bridge
medium.  The \HI\ contour maps of Mathewson \& Ford (1984) indicate an
\HI\ column density between 1$\times10^{20}$ and 5$\times10^{20}$ \cmmb\
while our pointed observations better constrain this value to within 1--2
$\times10^{20}$ \cmmb.

{\it B0637-752:\ } Bowen, Blades, \& Pettini (1995) detected two strong Mg~II
$\lambda\lambda$2796,2803 absorption components toward this quasar
at velocities of 250 \kms\ and 320 \kms\ with equivalent widths of 1.4
\AA\ and 1.1 \AA.  The \HI\ emission spectrum in Figure~2g shows a
component near 250 \kms\, and a secondary component near 290 \kms, but
no feature corresponding to the Mg~II absorption at 320 \kms.  Adopting
the mean LMC distance of 49.4 kpc used by Bowen \etal, this sightline
intersects the LMC at a radius of 6.7 kpc where we measure an
\HI\ columns density of 2.1$\times10^{19}$ \cmmb.  At this radius,
Dickey \etal\ (1994) find spin temperatures in excess of 200~K, or
possibly as high as 3000~K adopting their linear relation between
radius and spin temperature.  This is consistent with our lower limit
of $T_S>$380~K.

{\it B2353-685:\ }  We note the possible presence of an H~I absorption
feature with maximum optical depth $\tau=0.06$ near $v_\odot=$285
\kms\ in Figure~2h.  There is no detectable 21-cm emission at
corresponding velocities.  Optical or ultraviolet spectra toward this
QSO would be valuable to search for metal absorption lines which may
indicate the presence of a small, cold intergalactic cloud.

\subsection{Implications of the Measured Spin Temperatures}

In the Milky Way, requiring that the heating and cooling rates be equal
prescribes an equilibrium curve for density and temperature.  The
cooling rate increases as density squared since cooling processes
depend on collisions in the gas, (e.g., excitation of fine structure
lines) or recombination of electrons onto ions or charged grains.
Heating typically depends on density to the first power, since it is
proportional to the ionization rate due to cosmic rays or X-rays, or to
the photoelectron emission rate from grains.  The gas is forced to the
cool phase if the density is high, since the cooling rate will exceed
the heating rate, unless the gas is so cold that collisions seldom
excite the fine structure line of O~I.  For gas temperatures below 100~K 
the dominant coolants are C~II and C~I fine structure lines.   In the
Milky Way, for a given ISM pressure, there is a range of densities for
which the equilibrium is unstable.  Stable equilibria are possible only
if the gas density is below about 0.5 cm$^{-3}$ or above about 5
cm$^{-3}$, corresponding to temperatures above 6000~K and below 150~K
respectively (Field, Goldsmith \& Habing, 1969; reviewed by Wolfire 
\etal\ 1995).

Reducing the metallicity of the gas effects both the heating and
cooling processes, since in the cool phase the dominant heating process
is photoelectron emission from grains.  In the SMC the dust to gas
ratio is low by roughly an order of magnitude compared to the Milky Way
(Schwering 1989) and the abundance of C and O in the gas is low by a
factor of 8 (Pagel \etal\ 1978).  Wolfire et al. (1995) consider how
changing the metallicity of the medium changes the equilibrium curve of
heating-cooling balance.  They find that for an increasingly metal-poor
medium where the dust-to-gas ratio varies as the metallicity, the loss
of coolants is more important than the loss of grains for heating.  The
warm phase becomes more dominant, in the sense that the minimum
pressure for which the cool phase is possible increases.  The maximum
pressure for which the warm phase is possible also increases.

Spin temperatures of 20 to 50~K have been measured for HI clouds in the
SMC by Dickey et al.  (1998), thus the temperatures measured for the
Bridge gas here are similar to these SMC values.  In the Magellanic
Bridge the pressure must be very low on average, since there is no
steep gradient in the gravitational potential to confine gas at high
pressures, but the presence of cool HI clouds shows that some regions
must be at high densities.  Assuming a metallicity $Z=0.15 Z_\odot$
which is typical of nebulae in the SMC (Pagel \etal\ 1978), the minimum
density which allows a cool phase is about 10 cm$^{-3}$
(Wolfire \etal\ 1995; their Figure~6).  For
temperatures as low as 25 to 50~K, the density must be considerably
higher, probably above 100 cm$^{-3}$, if the gas is in thermal
equilibrium.  This implies a pressure of $3 \times 10^{3}$ cm$^{-3}$~K
for a temperature of 30~K.  The question is how to confine clouds with
this pressure.  It may be that this part of the Bridge is a region
where streamlines cross, meaning that the interaction between the
Clouds, and between the Clouds and the Milky Way, has led to
gravitational focusing of the gas trajectories in the Bridge region.
There may be a shock in the low density Bridge gas, with these cool
clouds tracing the post-shock material.

\subsection{Implications for Star Formation in the Magellanic Bridge}

The  dynamical structure  of the Magellanic Clouds is complex, probably
due to previous close interactions with each other, and with the Milky
Way (Mathewson \& Ford 1984; Murai \& Fujimoto 1980).  At least two
separate velocity components comprise the Magellanic system, and the
SMC has a large depth along the line of sight, perhaps 20 kpc,
suggesting that it is in the process of being disrupted (Wayte 1990;
Mathewson, Ford, \& Visvanathan 1986, 1988).   The bridge of neutral
hydrogen connecting the LMC and SMC 
(Hindman, Kerr, \& McGee 1963; Putman \etal\ 1998) 
is probably material that has been
tidally stripped from one of the galaxies, but its origin,
thermodynamic state, and metallicity are poorly known.

The Magellanic Stream exhibits both diffuse H$\alpha$ emission
(Johnson, Meaburn, \& Osman 1982; Meaburn 1986) and associations of
young blue stars in the eastern wing (Shapley 1940;
$\alpha=2^h~15^m,~\delta=-74\deg$) of the SMC and in the Bridge between
the Clouds (Courtes \etal\ 1995;   Grondin, Demers, \& Kunkel 1992;
Demers \etal\ 1991; Grondin \etal\ 1990; Irwin, Demers, \& Kunkel
1990).  Some of these clusters contain early B type stars with main
sequence lifetimes as short as 20 Myr.  They are located 5\degr
($\sim$5 kpc) from the dominant star-forming regions in the 
SMC, leading to speculation about their origin.  Have these young stars
migrated from regions of more vigorous star formation or have they
formed {\it in situ} from gas comprising the Bridge?  In order to
travel 5 kpc in 20 Myr, these stars must have extremely large peculiar
velocities in excess of 250 \kms.  An alternative interpretation is
that new stars, including massive stars, are currently forming in the
Magellanic Bridge.  The correlation between \HI\ column density and the
density of young OB associations in the Bridge (Battinelli \& Demers
1992) is consistent with this later hypothesis.  In all known
environments, star formation is observed to occur within molecular
clouds or cloud complexes, but there is presently no direct evidence of
star-forming gas clouds in the Bridge.

In lieu of millimeter-wave CO data which directly reveals the presence
of molecular gas, \HI\ 21-cm absorption traces the cold (30--40~K)
atomic medium which is associated with molecular clouds in the Galaxy
where the ratio of atomic to molecular column density,
$N_{HI}/N_{H_2}$, is $\sim0.02-0.2$ (Despois \& Baudry 1985).    The
detection of such cold cloud conditions even in an area of relatively
low \HI\ column density toward B0312-770 ($1\times10^{20}$ \cmmb)
strongly suggests the presence of atomic or molecular condensations
which could harbor star formation.   The B0312-770 sightline passes
just south of the Bridge regions searched by Bottinelli \& Demers
(1992) and Irwin \etal\ (1990) for OB associations, so we cannot test
whether this particular ISM feature corresponds to a stellar
association.

\section{Conclusions}

We present \HI\ 21-cm emission and absorption measurements toward 8
sightlines in the periphery and Bridge region of the Magellanic
Clouds.  Toward two of the background radio sources, we detect
measurable 21-cm absorption which indicates the presence of cold
atomic gas with spin temperatures between 20~K and 40~K.  The presence
of cold atomic condensations in the relatively low column density
outlying regions of the Clouds is direct evidence for the raw material
of star formation.   If the young OB associations observed in the
Magellanic Bridge region formed {\it in situ}, then our nearest neighbor
galaxies offer a laboratory for studying the star formation process in
the tidal tails and streamers produced during galactic interactions.
This result underscores the need for a more thorough study of the
thermodynamic properties of the gas and of the stellar populations in the
Magellanic Bridge.

\acknowledgements  We thank Mary Putman for producing
the 21-cm H~I map of the Magellanic System for Figure~1
and Snezana Stanimirovic for help with the data collection and
reduction.
H.~A.~K. thanks the Guillarmo Haro International
Program for Advanced Studies in Astrophysics at the Instituto National
de Astrofisica Optica y Electronica (INAOE) in Puebla, Mexico for
hospitality during the 1998 summer workshop on galaxy evolution during
which much of this paper was written. H.~A.~K. is grateful for
assistance from a NASA Graduate Student Researchers Program fellowship
and from NASA through grant \#HF-01094.01-97A awarded by the Space Telescope Science
Institute which is operated by the Association of Universities for
Research in Astronomy, Inc. for NASA under contract NAS 5-26555.  This
research has made use of the NASA/IPAC Extragalactic Database (NED)
which is operated by the Jet Propulsion Laboratory, California
Institute of Technology, under contract with the National Aeronautics
and Space Administration.

\clearpage

\begin{figure}
\centerline{\psfig{file=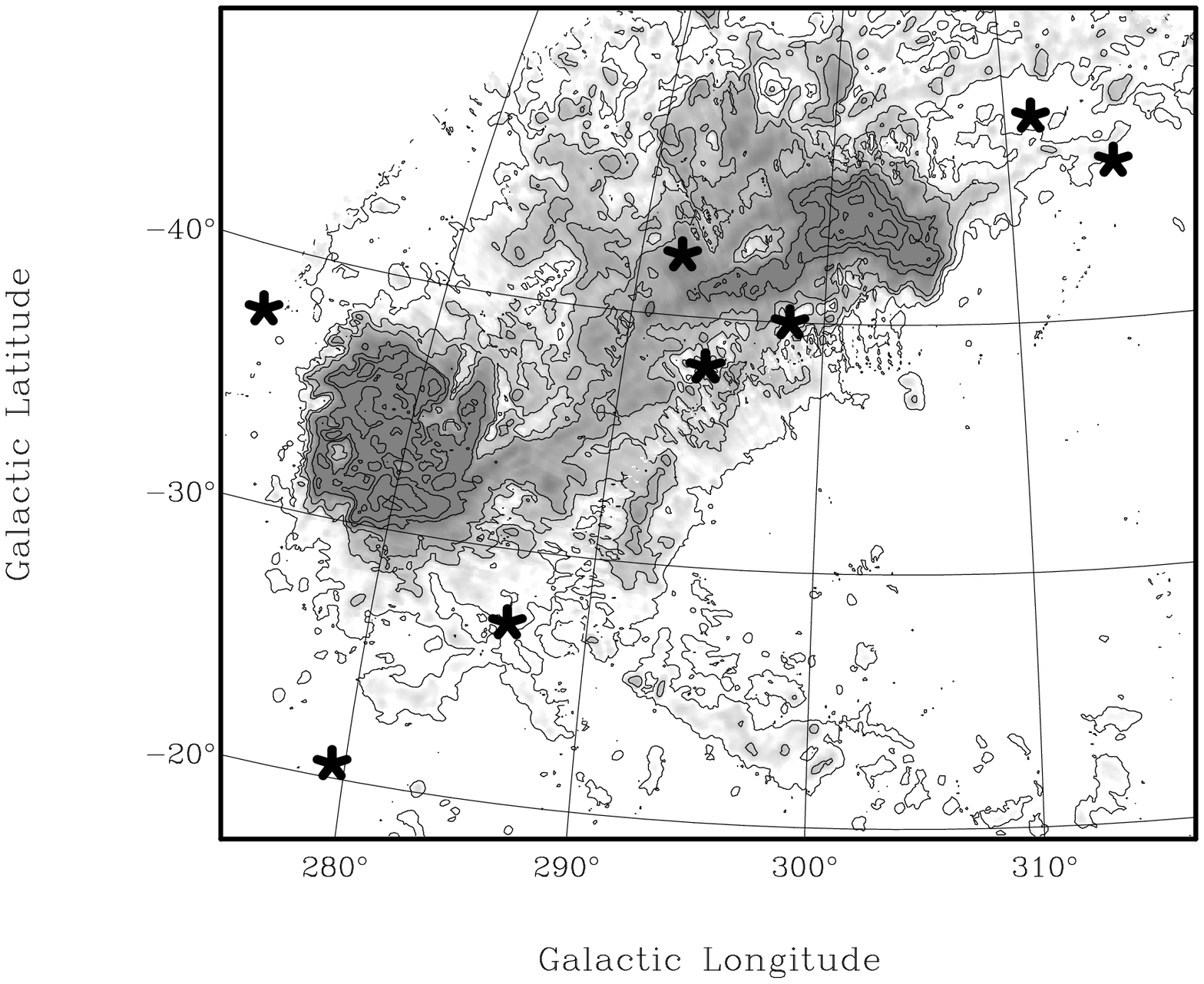,width=6.5in,angle=0}}
\figcaption[f1.ps] {H~I peak brightness temperature 
map of the Magellanic Clouds from
Putman \etal\ (1998).  Contours mark the 0.1 K (5$\sigma$), 0.8, 2, 8,
16, 32, 64, and 128 K peak brightness.  Asterisks mark the
lines of sight toward background radio sources listed in Table~1.  Two
radio sources lie behind the Magellanic Bridge, while the others lie on
the extreme periphery of the Clouds along lines of sight with low
\HI\ column density ($N_{HI}<4\times10^{19}$ \cmmb).  \label{Mclouds} }
\end{figure}

\begin{figure}
\centerline{\psfig{file=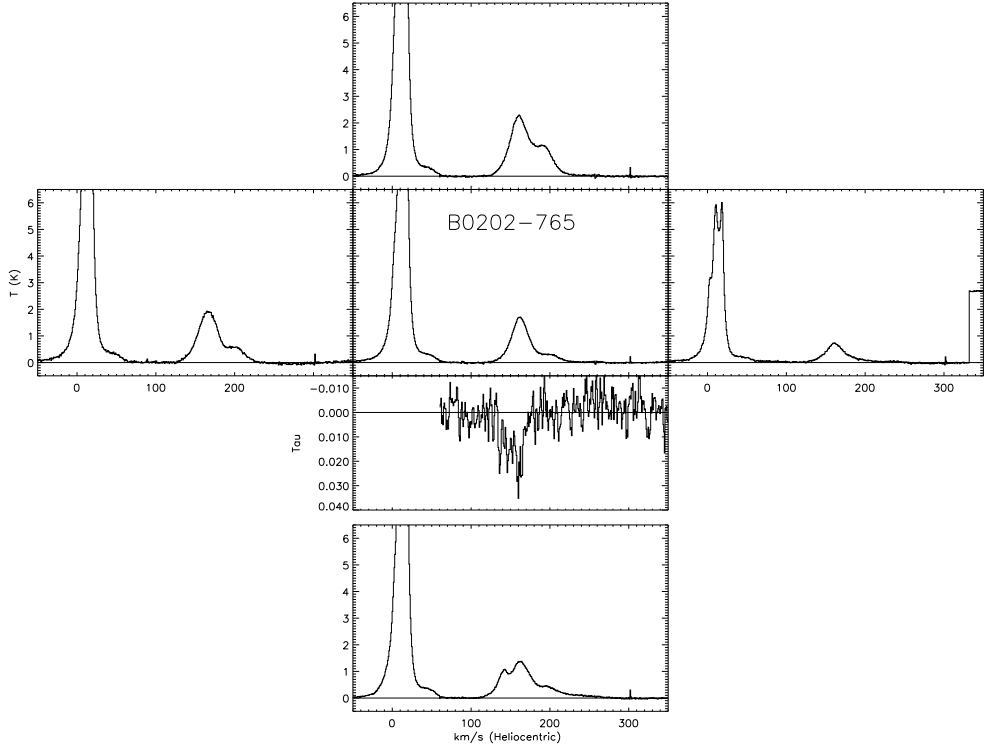,width=5.5in,angle=0}}
\figcaption[f2.ps] {a: 21-cm emission and absorption spectra along
8 lines of sight toward the Magellanic Bridge and around the periphery
of the LMC and SMC.  The central panel in each figure shows the
\HI\ emission spectrum from the Parkes 64 m antenna toward the
background radio source.  Absorption spectra against the background
radio sources obtained with the Australia Telescope Compact Array appear
just below the central panel.  Flanking emission spectra show the
\HI\ profiles offset by 14\arcmin\ to either side of
the primary sightline.  The channel width of the absorption and
emission data is 3.9 kHz (0.82 \kms).  The absorption spectra have been
smoothed with a 3-pixel boxcar for display purposes.
\label{spectra} }
\end{figure}

\begin{figure}
\centerline{\psfig{file=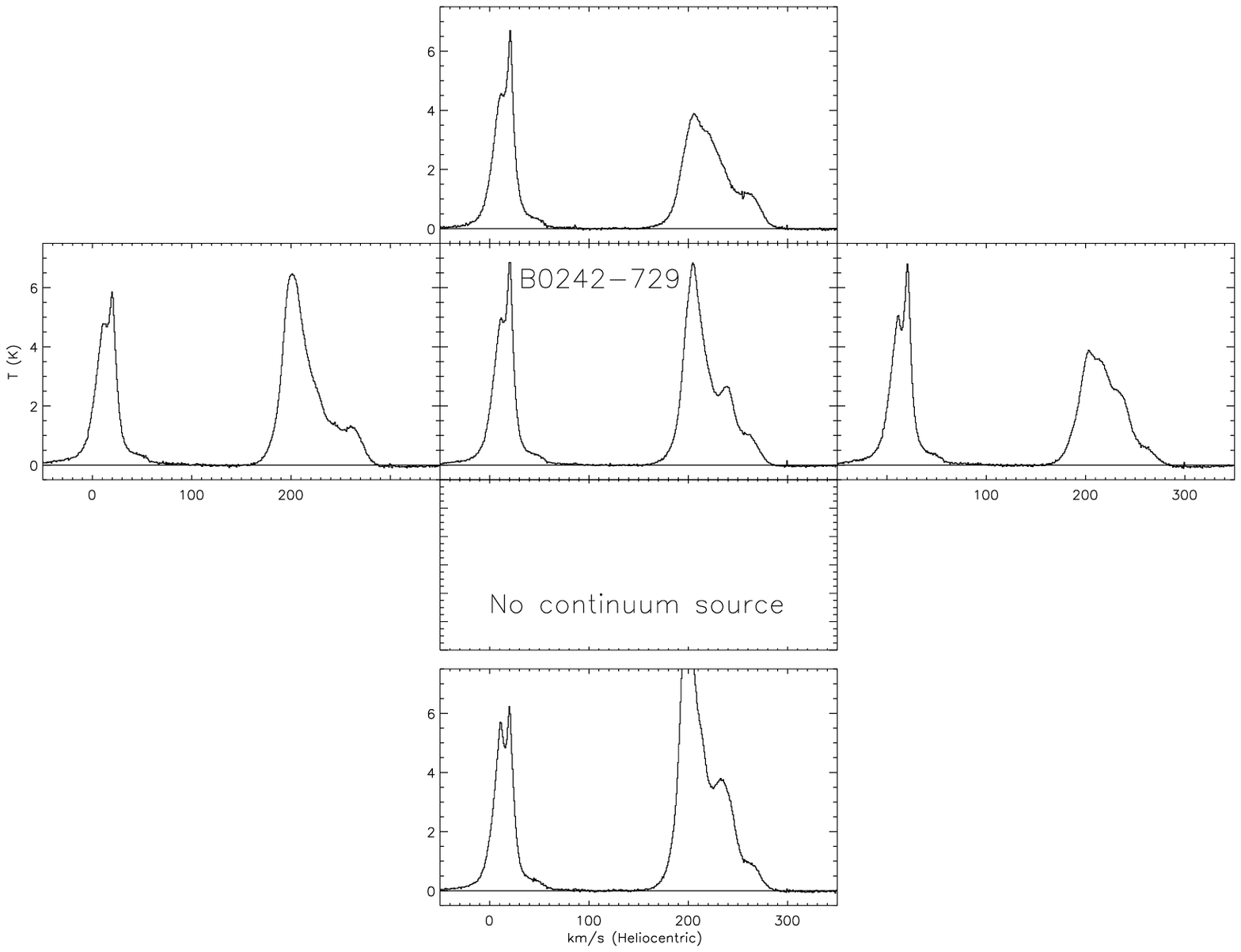,width=5.5in,angle=0}}
{Fig. 2b}
\end{figure}

\begin{figure}
\centerline{\psfig{file=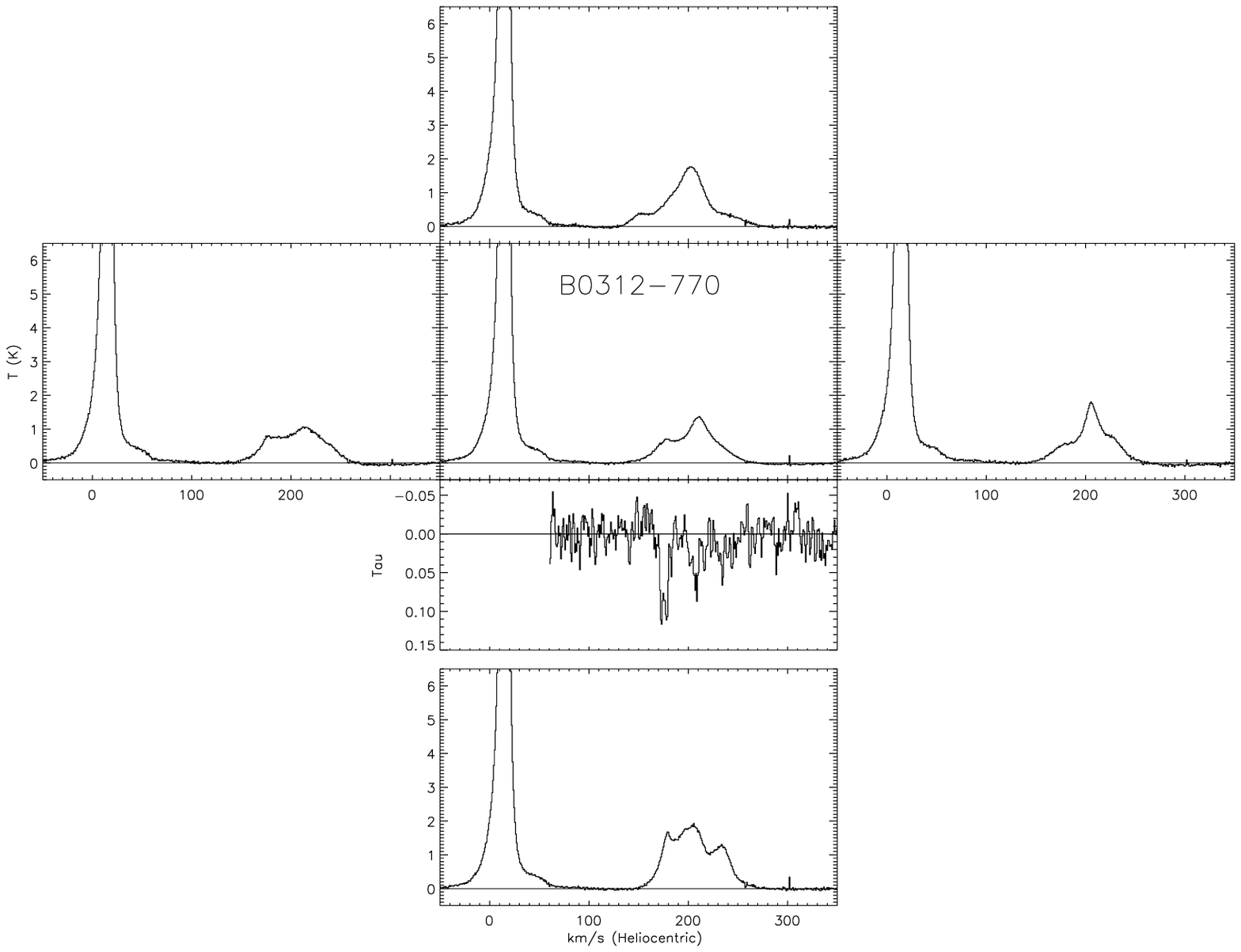,width=5.5in,angle=0}}
{Fig. 2c}
\end{figure}

\begin{figure}
\centerline{\psfig{file=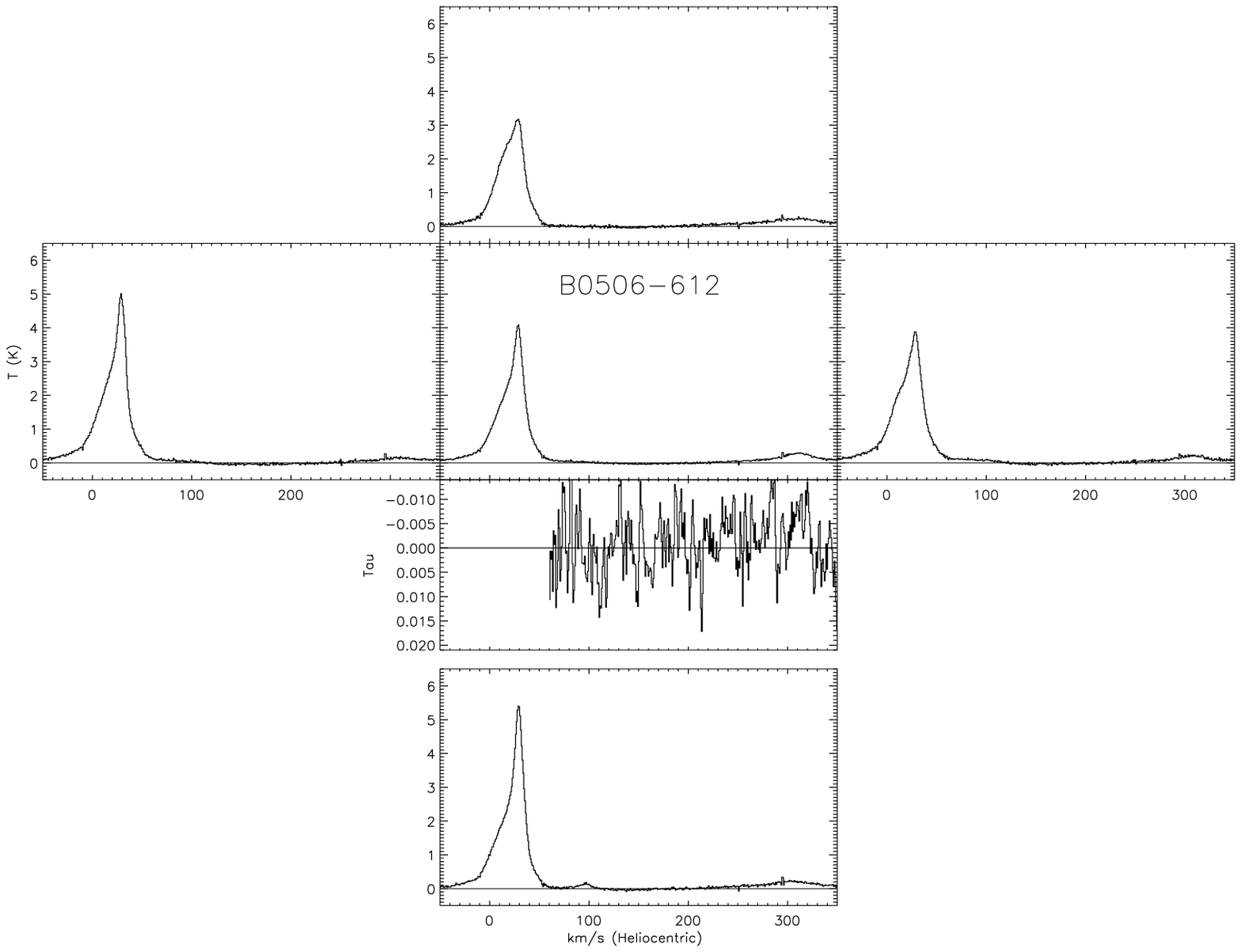,width=5.5in,angle=0}}
{Fig. 2d}
\end{figure}

\begin{figure}
\centerline{\psfig{file=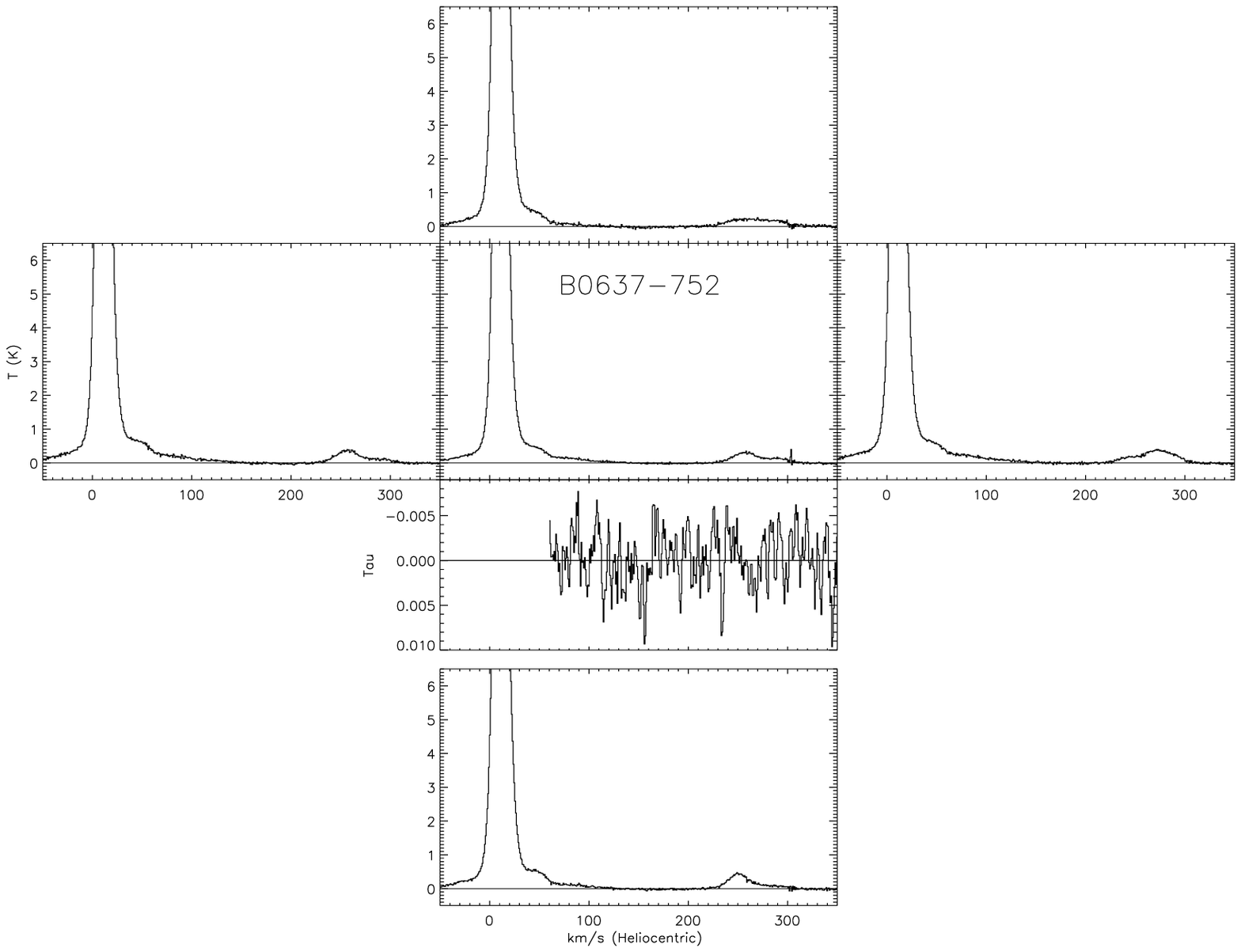,width=5.5in,angle=0}}
{Fig. 2e}
\end{figure}

\begin{figure}
\centerline{\psfig{file=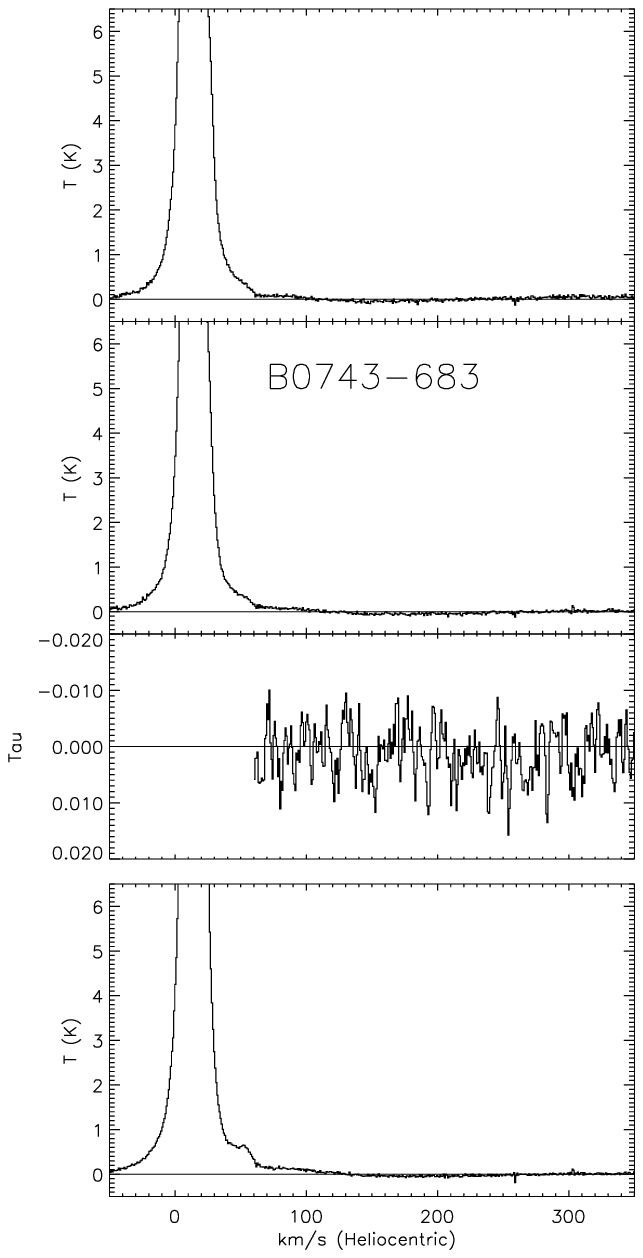,width=5.5in,angle=0}}
{Fig. 2f}
\end{figure}

\begin{figure}
\centerline{\psfig{file=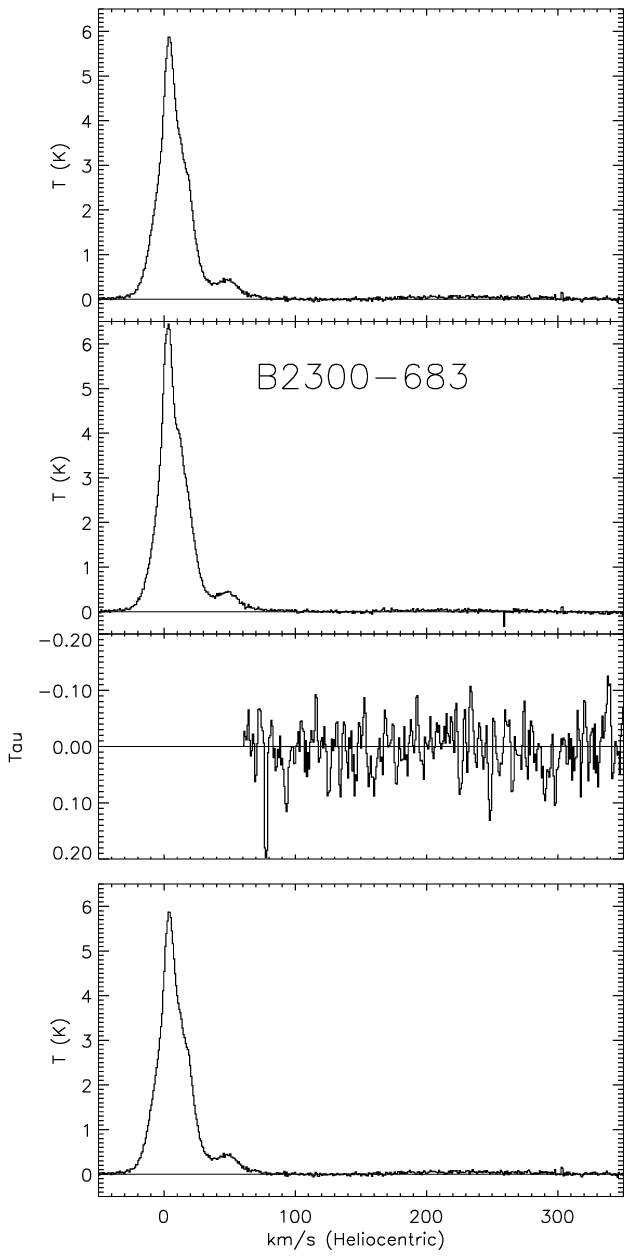,width=5.5in,angle=0}}
{Fig. 2g}
\end{figure}

\begin{figure}
\centerline{\psfig{file=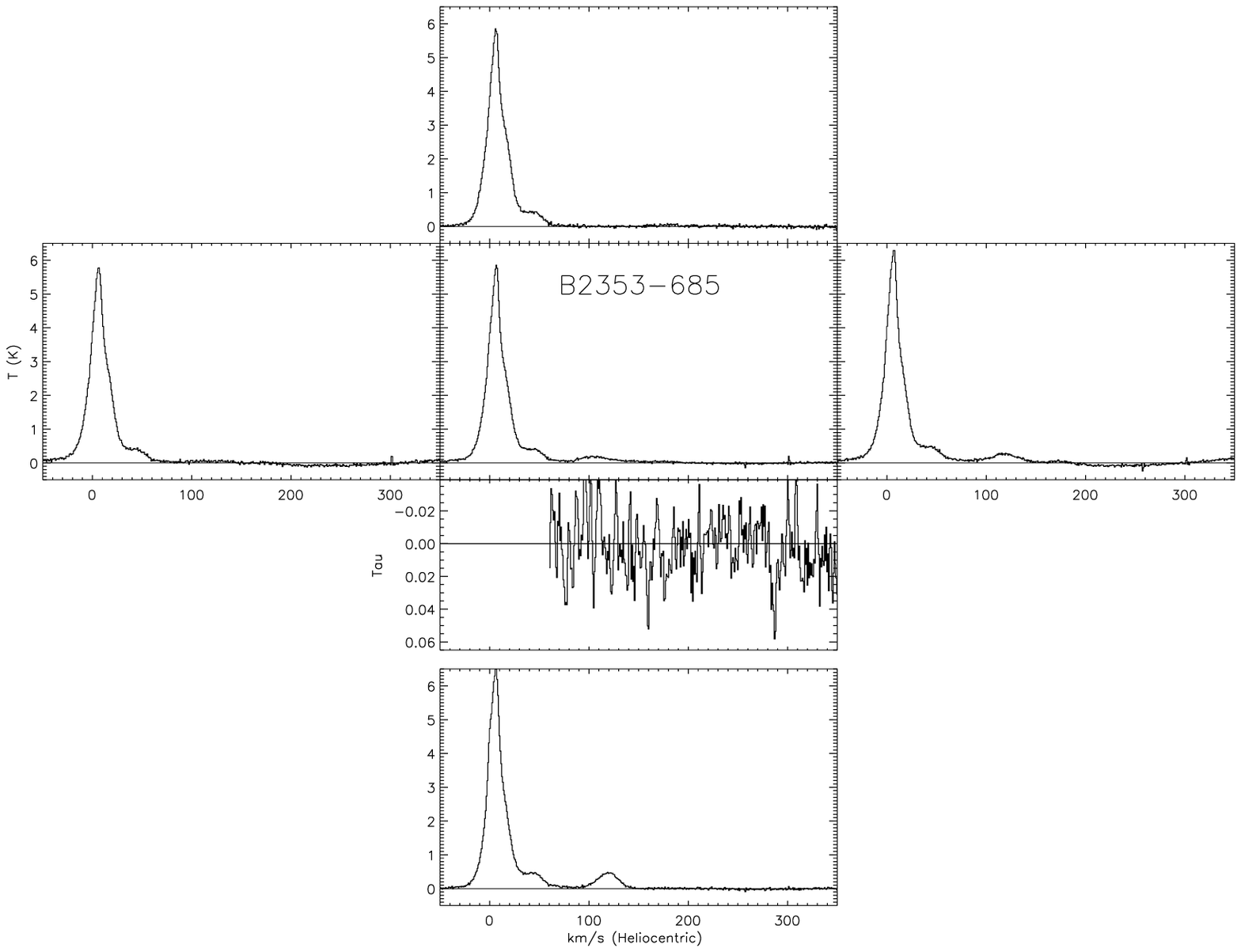,width=5.5in,angle=0}}
{Fig. 2h}
\end{figure}

\begin{figure}
\centerline{\psfig{file=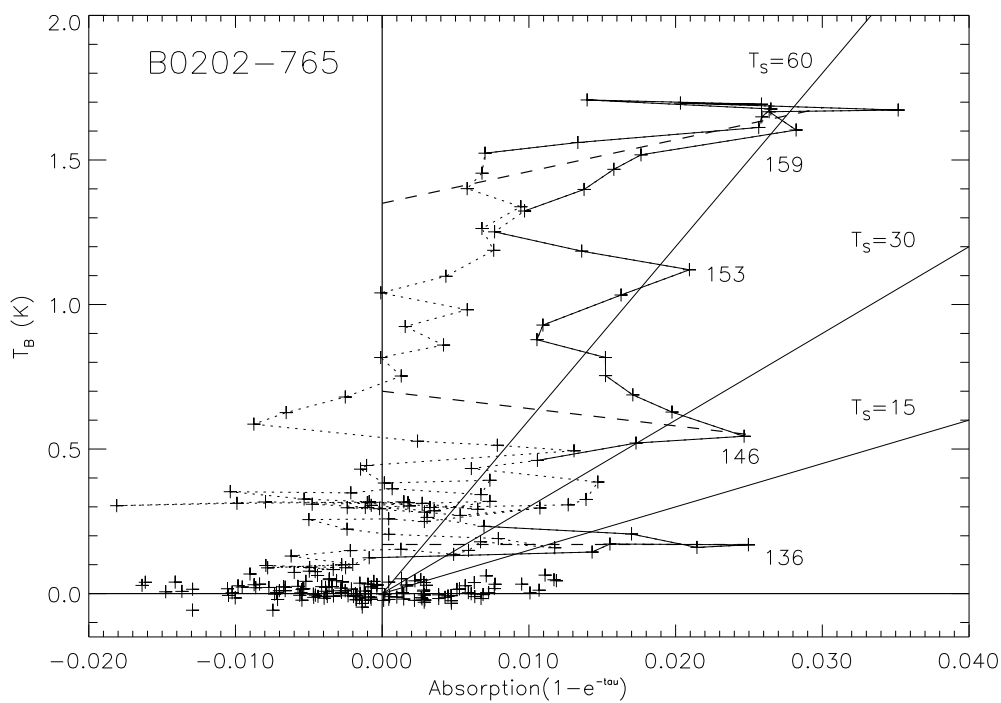,width=5.5in,angle=0}}
\figcaption[f3.ps] {The \HI\ emission brightness temperature, $T_B$,
versus absorption (1-$e^{-\tau}$) for each 0.82 \kms\ velocity channel
(crosses) in the B0202-765 spectra.  Velocity intervals with measurable
absorption appear connected by solid lines, while regions of
measurable emission appear connected by dotted lines.  Numbers adjacent
to the symbols give the heliocentric velocity of the channel.  Solid
lines from the origin illustrate the locii of constant spin
temperature, $T_s$.  Dashed line segments show the
fit to specific absorption features and are used to derive the spin
temperature of the cold atomic component.
\label{0202diag} }
\end{figure}

\begin{figure}
\centerline{\psfig{file=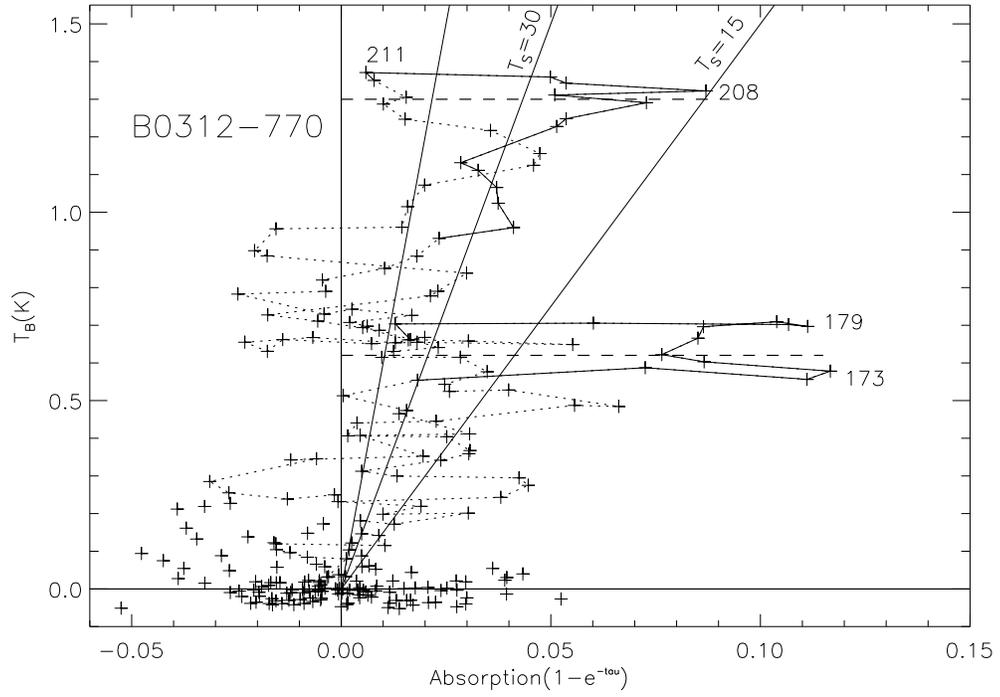,width=5.5in,angle=0}}
\figcaption[f4.ps] {Same as Figure~3 for
the sightline toward B0312-770.  
\label{0312diag} }
\end{figure}

\end{document}


\begin{planotable}{cccccc}
\tablewidth{0pt}
\tablecaption{Log of Parkes and ATCA 21-cm Observations \label{Obs} }
\tablehead{
\colhead{Object} & 
\colhead{RA (2000)\tablenotemark{a}}   & 
\colhead{DEC (2000)\tablenotemark{a}}      &
\colhead{Flux } &
\colhead{Flux} & 
\colhead{B mag.  } \\ [.2ex]
\colhead{} &
\colhead{} &
\colhead{} &
\colhead{(mJy)\tablenotemark{b} }& 
\colhead{(mJy)\tablenotemark{c}} &
\colhead{} }
\startdata
B2300-683 & 23 03 44.18 & -68 07 45.5 & 320 & ...  & 16.4 \nl 
B2353-686 & 23 56 05.93 & -68 18 41.9 & 960 & 1180 & 17.9 \nl
B0202-765 & 02 02 13.08 & -76 20 05.3 & 2340& 2440 & 16.0 \nl 
B0242-729 & 02 43 09.59 & -72 16 48.9 & $<$66\tablenotemark{d} & ... & 16. \nl 
B0312-770 & 03 11 54.75 & -76 51 51.6 & 610 & 620  & 17.2 \nl
B0506-612 & 05 06 44.10 & -61 09 40.9 & 2010& 2000 & 17.9 \nl
B0637-752 & 06 37 46.53 & -75 16 17.1 & 3730& 4940 & 17.2 \nl 
B0743-673 & 07 43 32.35 & -67 26 25.6 & 5320& 5300 & 17.0 \nl 
\enddata 
\tablenotetext{a}{Optical  position}
\tablenotetext{b}{Measured ATCA flux density for the 
radio source mapped with a synthesized beamsize of $\sim$40\arcsec}
\tablenotetext{c}{Flux from Parkes PKSCAT90 catalog (Otrupcek \& Wright 1991)}
\tablenotetext{d}{3$\sigma$ upper limit}
\end{planotable}
\clearpage

\begin{planotable}{lccrrccrrrr}
\tablewidth{0pt}
\tablecaption{21-cm Emission and Absorption Line Profile Data \label{Obs} }
\tablehead{
\colhead{Object} & 
\colhead{Emission} &
\colhead{} &
\colhead{} &
\colhead{} &
\colhead{} &
\colhead{Absorption} &
\colhead{} &
\colhead{} &
\colhead{} &
\colhead{}  \\ [.2ex]
\colhead{} &
\colhead{$\int T_Bdv$\tablenotemark{a}}   & 
\colhead{$N_{HI}$}   & 
\colhead{$V_{\odot}$}  &
\colhead{FWHM} & 
\colhead{Peak} &  
\colhead{EW} &
\colhead{$V_{\odot}$} &
\colhead{FWHM} &
\colhead{$\tau_{max}$} &
\colhead{$\tau_{RMS}\tablenotemark{b}$}  \\ [.2ex]
\colhead{} &  
\colhead{K km s$^{-1}$} &  
\colhead{$\times10^{19}~cm^{-2}$} &  
\colhead{\kms}  &
\colhead{\kms}  &
\colhead{K}     &
\colhead{\kms} &
\colhead{\kms} &
\colhead{\kms} &
\colhead{} &
\colhead{}  \\ [.2ex]
\colhead{(1)} &
\colhead{(2)} &
\colhead{(3)} &
\colhead{(4)} &
\colhead{(5)} &
\colhead{(6)} &
\colhead{(7)} &
\colhead{(8)} &
\colhead{(9)} &
\colhead{(10)} & 
\colhead{(11)} }
\startdata
B2300-683  & $<$0.6\tablenotemark{a} & $<$0.11\tablenotemark{a}        & ...  & ...  & $<$0.05 & $<$0.45\tablenotemark{b}  & ...  & ...  & ...   & 0.076  \nl
B2300-683n & $<$0.6\tablenotemark{a} & $<$0.11\tablenotemark{a}        & ...  & ...  & $<$0.05 & ...             & ...  & ...  & ...   & ...    \nl
B2300-683s & $<$0.6\tablenotemark{a} & $<$0.11\tablenotemark{a}        & ...  & ...  & $<$0.05 & ...             & ...  & ...  & ...   & ...    \nl
B2353-685  & 8.08$\pm$0.15  &  1.47$\pm$0.03     & ...  & ...  & 0.21 & $<$0.15\tablenotemark{b} & ...  & ...  & ...   & 0.032  \nl
B2353-685n & 1.79$\pm$0.13  &  0.37$\pm$0.02     & ...  & ...  & 0.10 & ...             & ...  & ...  & ...   & ...    \nl
B2353-685s & 11.40$\pm$0.11 &  2.08$\pm$0.02     & ...  & ...  & 0.49 & ...             & ...  & ...  & ...   & ...    \nl
B2353-685e & 0.0$\pm$0.22  &  0.0$\pm$0.04     & ...  & ...  & 0.11 & ...             & ...  & ...  & ...   & ...    \nl
B2353-685w & 6.97$\pm$0.19 &  1.27$\pm$0.03     & ...  & ...  & 0.29 & ...             & ...  & ...  & ...   & ...    \nl
B0202-765  & 55.98$\pm$0.31 & 10.24$\pm$0.06     & ...  & ...  & 1.71 & 0.62$\pm$0.05   & ...  & ...  & 0.035 & 0.010  \nl 
\ \ comp 1      & ...            &  ...          & ...  & ...  & ...     & 0.07$\pm$0.01   & 137.4& 3.0  & 0.023 & 0.010  \nl 
\ \ comp 2      & ...            &  ...          & ...  & ...  & ...     & 0.19$\pm$0.03   & 147.6& 10.0 & 0.018 & 0.010  \nl 
\ \ comp 3      & 44.60$\pm$0.10 &  8.16$\pm$0.18& 161.8& 25.5 & 1.63    & 0.31$\pm$0.04   & 162.0& 11.1 & 0.025 & 0.010  \nl 
\ \ comp 4      & 9.95$\pm$0.07  &  1.82$\pm$0.12& 197.8& 31.2 & 0.30    & ...             & ...  & ...  & ...   & 0.010  \nl 
B0202-765n & 96.34$\pm$0.16 & 17.63$\pm$0.03& ...  & ...  & 2.30 & ...             & ...  & ...  & ...   & ...    \nl 
B0202-765s & 71.05$\pm$0.39 & 13.00$\pm$0.07& ...  & ...  & 1.38 & ...             & ...  & ...  & ...   & ...    \nl 
B0202-765e & 74.54$\pm$0.46 & 13.64$\pm$0.08& ...  & ...  & 1.93 & ...             & ...  & ...  & ...   & ...    \nl 
B0202-765w & 24.83$\pm$0.35 &  4.54$\pm$0.06& ...  & ...  & 0.74 & ...             & ...  & ...  & ...   & ...    \nl 
B0242-729  &251.74$\pm$0.50 & 46.07$\pm$0.09& ...  & ...  & 6.94    & ...              & ...  & ...  & ...   & ...    \nl
B0312-770  & 65.40$\pm$0.38 & 11.97$\pm$0.07& ...  & ...  & 1.37    & 2.05$\pm$0.26   & ...  & ...          & 0.120 & 0.040  \nl
\ \ comp 1      & 24.30$\pm$0.38 & 4.44$\pm$0.69 & 181.1& 33.5 & 0.68    & 1.08$\pm$0.07     & 175.8&  9.6  & 0.105 & 0.030  \nl
\ \ comp 2      & 22.17$\pm$0.37 & 4.05$\pm$0.68 & 209.6& 20.8 & 1.00    & 0.77$\pm$0.07     & 206.9&  13.2 & 0.055 & 0.030  \nl
\ \ comp 3      & 19.65$\pm$0.10 & 3.59$\pm$0.18 & 227.9& 34.9 & 0.53    & 0.43$\pm$0.07     & 234.6&  10.5 & 0.034 & 0.030  \nl
B0312-770n & 84.78$\pm$0.40 & 15.51$\pm$0.07& ...  & ...  & 1.77 & ...             & ...  & ...  & ...   & ...    \nl
B0312-770s &110.89$\pm$0.48 & 20.29$\pm$0.09& ...  & ...  & 1.93 & ...             & ...  & ...  & ...   & ...    \nl
B0312-770e & 66.38$\pm$0.44 & 12.14$\pm$0.08& ...  & ...  & 1.08 & ...             & ...  & ...  & ...   & ...    \nl
B0312-770w & 66.84$\pm$0.41 & 12.23$\pm$0.07& ...  & ...  & 1.81 & ...             & ...  & ...  & ...   & ...    \nl
B0506-612  & 19.01$\pm$0.27 &  3.47$\pm$0.05& ...  & ...  & 0.30 & $<$0.07\tablenotemark{b}  & ...  & ...  & ...   & 0.011  \nl
B0506-612n & 20.71$\pm$0.45 &  3.79$\pm$0.08& ...  & ...  & 0.34 & ...             & ...  & ...  & ...   & ...    \nl
B0506-612s & 17.51$\pm$0.52 &  3.20$\pm$0.09& ...  & ...  & 0.34 & ...             & ...  & ...  & ...   & ...    \nl
B0506-612e & 12.83$\pm$0.33 &  2.34$\pm$0.06& ...  & ...  & 0.27 & ...             & ...  & ...  & ...   & ...    \nl
B0506-612w & 14.27$\pm$0.29 &  2.28$\pm$0.06& ...  & ...  & 0.27 & ...             & ...  & ...  & ...   & ...    \nl
B0637-752  & 11.48$\pm$0.28 &  2.10$\pm$0.06& ...  & ...  & 0.35 & $<$0.03\tablenotemark{b} & ...  & ...  & ...   & 0.005  \nl 
B0637-752n & 13.36$\pm$0.32 &  2.44$\pm$0.06& ...  & ...  & 0.26 & ...             & ...  & ...  & ...   & ...    \nl 
B0637-752s & 12.76$\pm$0.29 &  2.33$\pm$0.06& ...  & ...  & 0.47 & ...             & ...  & ...  & ...   & ...    \nl 
B0637-752e & 12.74$\pm$0.46 &  2.33$\pm$0.08& ...  & ...  & 0.39 & ...             & ...  & ...  & ...   & ...    \nl 
B0637-752w & 16.59$\pm$0.26 &  3.03$\pm$0.05& ...  & ...  & 0.40 & ...             & ...  & ...  & ...   & ...    \nl 
B0743-683  & $<$0.6\tablenotemark{a} & $<$0.11\tablenotemark{a}       & ...  & ...  & $<$0.05 & $<$0.05\tablenotemark{b}  & ...  & ...  & ...   & 0.008  \nl 
B0743-683n & $<$0.6\tablenotemark{a}& $<$0.11\tablenotemark{a}       & ...  & ...  & $<$0.05 & ...             & ...  & ...  & ...   & ...    \nl 
B0743-683s & $<$0.6\tablenotemark{a} & $<$0.11\tablenotemark{a}       & ...  & ...  & $<$0.05 & ...             & ...  & ...  & ...   & ...    \nl 
\enddata 
\tablenotetext{1}{Object Name.  Where multiple velocity components can be defined,
they are  listed seperately.}
\tablenotetext{2}{21-cm emission line integral from 100 \kms\ --- 350 \kms\ or the integral of the 
	best-fit Gaussian profile in the case of individual components.}   
\tablenotetext{3}{\HI\ column density derived from column 1 assuming 
	$N_{HI}~(cm^{-2})=1.823\times10^{18} \int
	T_B~dv~$ for an optically thin medium.}
\tablenotetext{4}{The center velocity of best-fit Gaussian components to the emission spectrum in the
	heliocentric reference frame.} 
\tablenotetext{5}{The full width half max of the best-fit Gaussian components in \kms.}
\tablenotetext{6}{The peak 21-cm brightness temperature from the Parkes 64 m spectra. 
	The listed value corresponds to the global maximum $T_B$ or
	the maximum of the Gaussian fit in the case of individual components.}
\tablenotetext{7}{The absorption equivalent width. The listed value corresponds to the global line integral,
$\int\tau~dv$, or the integral of the best fit Gaussian profile in the case
of individual components.}
\tablenotetext{8}{The center velocity of the best-fit Gaussian profile for individual
absorption components.}
\tablenotetext{9}{The full width at half depth of the best-fit Gaussian profile for
	absorption components.}
\tablenotetext{10}{Maximum optical depth.  The listed value corresponds to the global
maximum in the optical depth, or the depth of the Gaussian fit in
the case of individual components.  }  
\tablenotetext{11}{Channel-to-channel RMS noise in the raw, 0.82 \kms\ unsmoothed spectra.  For display purposes,
	the absorption spectra are smoothed with a 3-pixel boxcar in the Figures where the RMS appears a factor
	$\sim$2 lower. }
\tablenotetext{a}{3$\sigma$ upper limit}
\tablenotetext{b}{3$\sigma$ upper limit for an assumed line width of 3.3 \kms\ (4 channels) }
\end{planotable}
\clearpage

\begin{planotable}{crr}
\tablewidth{0pt}
\tablecaption{Derived Properties of Absorption-Line Sources \label{Obs} }
\tablehead{
\colhead{Object} & 
\colhead{$<T_s>$}   & 
\colhead{$T_s(v_0)$}      \\ [.2ex]
\colhead{} &
\colhead{K} &
\colhead{K}  \\ [.2ex]
\colhead{(1)} &
\colhead{(2)} &
\colhead{(3)} }
\startdata
B2353-685   & $>$42      & $>$7   \nl
B0202-765   & 87$\pm$7   & ...    \nl
 \ \ comp 1 & ...        & 8      \nl
 \ \ comp 2 & ...        & 38     \nl
 \ \ comp 3 & 143$\pm$19 & 68     \nl 
B0312-770   & 32$\pm$4   & ...    \nl
 \ \ comp 1 & 22$\pm$2   & 6      \nl
 \ \ comp 2 & 29$\pm$3   & 38     \nl
 \ \ comp 3 & 46$\pm$7   & 14     \nl
B0506-612    & $>$270     & $>$27 \nl
B0637-752   & $>$380     & $>$70  \nl
\enddata 
\tablenotetext{1}{Object Name.  Fits to individual components are listed seperately.}
\tablenotetext{2}{The ensemble average spin temperature.
	computed from $\int{T_B}dv/EW_{abs}$ using the data from Table~2.}
\tablenotetext{3}{The spin temperature, $T_S$, computed in the traditional fashion, 
	for optically-thin clouds, $T_s(v_0)=T_B(v_0)/(1-e^{-\tau(v_0)})=T_B(v_0)/\tau(v_0)$ when $\tau\ll{1}$.  
	$T_B$ is the brightness temperature at velocity $v_0$ in
	the emission spectrum, and $\tau$ the optical depth at the same velocity. 
	We take $v_0$ to be the center velocity in the case of individual Gaussian
	component.}
\end{planotable}
\clearpage